\documentclass{article}
\usepackage{amsmath,amssymb}

\def\bR{{\bf R}}

\def\bZ{{\bf Z}}
\def\Im{\mathop{\rm Im}\nolimits}
\def\Re{\mathop{\rm Re}\nolimits}

\def\sh{\mathop{\rm sh}\nolimits}

\def\CC{{\cal C}}

\def\EE{{\cal E}}

\def\TT{{\cal T}}

\def\k{{\kappa}}

\def\wh{\widehat}

\def\interior#1{\setbox1=\hbox{$#1$}\rlap{$#1$}\kern0.4\wd1\raise1.1\ht1%
\hbox{$\scriptstyle \circ$}}

\def\boxit#1#2{\setbox1=\hbox{\kern#1{#2}\kern#1}%
\dimen1=\ht1 \advance \dimen1 by #1 \dimen2=\dp1 \advance \dimen2 by
#1
\setbox1=\hbox{\vrule height\dimen1 depth\dimen2\box1\vrule}%
\setbox1=\vbox{\hrule\box1\hrule}%
\advance \dimen1 by .4pt \ht1=\dimen1 \advance \dimen2 by .4pt
\dp1=\dimen2 \box1\relax}
\def\endprf{\raise .5ex\hbox{\boxit{2pt}{\ }}}

\def\half{{\scriptstyle{1 \over 2}}}

\def\ifundefined#1{\expandafter\ifx\csname#1\endcsname\relax}

\def\beq{\begin{equation}}
\def\endq{\end{equation}}
\def\beqa{\begin{eqnarray}}
\def\endqa{\end{eqnarray}}

\begin{document}

\title{Infrared surprises in the de Sitter universe\footnote{Dedicated to Mario Novello for his birthday}}
\author{Ugo Moschella\\Universit\`a dell'Insubria Como Italia \\ and INFN sezione di Milano, Italia}

\maketitle
\begin{abstract}
We describe a few unexpected features of de Sitter Quantum Field Theory that have no Minkowskian counterparts.
These phenomena show that even when the cosmological constant is tiny
a Minkowskian way of fast thinking about de Sitter can lead to mistakes and that de Sitter QFT is essentially
different from standard relativistic (Minkowskian) QFT. 
The results reported in this paper are in collaboration with {Jacques Bros} and {Henri Epstein}.
\end{abstract}

\section{Introduction}
The astronomical observations  of the last fifteen years have led to the surprising conclusion that the recent universe is dominated by an almost spatially homogeneous exotic form of energy density to which there corresponds a negative pressure. Such negative pressure acts repulsively at large scales, opposing itself to the gravitational attraction. This effect may explain the accelerated expansion of the universe and may account for an important part of the missing mass. It has become customary to characterize such energy density by the term dark.

The simplest and best known candidate for the dark energy is the cosmological constant. It was Einstein himself who introduced a constant term in the equations for the gravitational field as a mechanism to obtain static cosmological solutions. However, this possibility was immediately set aside because the static solution so obtained was unstable and, even more, because the observations performed shortly after by Edwin Hubble pointed towards a non static expanding universe. If there is no quasi-static world then away with the cosmological constant (postcard from Eistein to Hermann Weyl -- 1923) and the cosmological constant was to be downgraded to a mere mathematical curiosity.

This consolidated belief was shattered and quickly abandoned in 1998, following the discovery that the expansion of the universe in the present epoch is accelerated. This discovery was at first based on measurements of the distance-red shift relation of distant supernovae, but it has been subsequently confirmed and strengthened by independent observations.

In the above context the de Sitter geometry, which is the homogeneous and isotropic solution of the vacuum Einstein equations with cosmological term, appears to take the role of reference geometry of the universe. In other words, it is the de Sitter geometry, and not the Minkowski one, which would be the geometry of empty spacetime (namely of spacetime deprived of its matter and radiation content). In addition, if the description provided by the standard $\Lambda$CDM model is correct, the
remaining energy components must in the future progressively
thin out and eventually vanish thus letting the cosmological
constant term alone survive. Therefore, the de Sitter geometry is the one to which the geometry of the universe approaches asymptotically. These considerations show the actuality and the importance that the de Sitter geometry acquires in present day cosmology, in addition to the traditional role which it plays in the context of inflationary models; the de Sitter geometry takes a role in contemporary cosmology that is in a way more relevant than the one played by the flat Minkowski geometry.

\section{de Sitter world}
The easiest way to visualise a $d$-dimensional de Sitter universe with radius $R$
is to think of it as the one-sheeted hyperboloid
\begin{equation}
X_d = \{x\in {\bf R}^{d+1}, \ \  x^2 = x\cdot x =  -R^2\}
\label{desittermanifold}
\end{equation}
embedded in a $(d+1)$-dimensional Minkowski space time.
The relation between the radius $R$, the cosmological constant $\Lambda$ and the constant scalar curvature $\kappa$ are given by
$$ \Lambda = {\frac{(d-1)(d-2)}{2R^2}}\, ,  \ \ \ \ \kappa = \frac{d(d-1)}{R^2}. $$
There is a global causal ordering relation on $X_{d}$ induced
by that of ${{\bf R}}^{d+1}$.  Let
\begin{equation}
{V^{+}} = \left\{x\in {{\bf R}}^{d+1}:
{x_{0}}>\sqrt{{x_{1}}^{2}+\ldots +{x_{d}}^{2}}\right\}
\end{equation}
be the forward cone in the ambient spacetime.
Then  two events $x,{x'}\in X_d$ are causally ordered ($x$ is in the future of $x'$) if and only if  $x-{x'} \in {V^{+}}$ in the ambient spacetime sense.

The de Sitter relativity group coincides with the Lorentz
group of the ambient spacetime. There are no commutative
translations on the de Sitter manifold and this very fact is the source
of considerable technical difficulties in the study (and the applications!) of de
Sitter Quantum Field Theory.

Next, we
consider the de Sitter Klein-Gordon equation for general complex values
of the squared mass $m^2$:
\begin{equation}
\square \phi + m^2 \phi = 0.
\end{equation}
This equation may be solved in several (equivalent, though formally different)
ways by separating the variables in one or another
of the many special coordinate systems that the maximal symmetry
of the de Sitter universe allows to set up.
However, there is also the possibility to avoid at this stage the use of any particular coordinate
system and work just with the embedding (\ref{desittermanifold}). For this purpose a key role is played by the forward asymptotic lightcone
\begin{equation}
C^+ = \{\xi\in {\bf R}^{d+1}, \ \  \xi^2 = \xi \cdot \xi = 0, \ \  \xi^0>0\}
\label{desittermanifold}
\end{equation}
 which is the boundary of the forward cone $V^+$. Forward-pointing lightlike vectors $\xi \in C^+$
 may be used to parameterize plane wave solutions of the Klein-Gordon Equation as follows:
\begin{equation}
\psi_\lambda(x, \xi) =  (\xi \cdot x)^\lambda.
\label{planewaves}
\end{equation}
A straightforward computation shows that
\begin{equation}
\square \psi_\lambda(x, \xi) -\frac{1}{R^2} \lambda(\lambda+d-1) \psi_\lambda(x, \xi)= 0
\end{equation}
and therefore the complex dimensionless parameter $\lambda$ is related to the squared mass by
\begin{equation}
m^2 =  -\frac{1}{R^2} \lambda(\lambda+d-1).
\label{mass}
\end{equation}

Plane waves are the quantum analogue of classical geodesics. Freely moving  Minkowskian particles,
as well as plane waves $\exp (ip\cdot x)$
are labeled by $d$-momenta $p$ satisfying  the mass-shell condition $p^2 = m^2$.
We may of course rewrite those waves in by separating the  momentum direction and the mass as follows:
\begin{equation} \exp (i m \hat p\cdot x), \ \ \ \ \hat  p ^2 = 1.\label{wm}
\end{equation}
Similarly the waves (\ref{planewaves}) are labeled by
the asymptotic direction  $\xi$ of a de Sitter freely moving particle and by the parameter $\lambda$ which is
related to the squared mass by Eq. (\ref{mass}).
The asymptotic cone  $C^+$ can then be physically
interpreted as the space of (asymptotic) momenta directions in close analogy with Eq. (\ref{wm}).

The complex squared mass $m^2 (\lambda)$ is invariant under the involution
\begin{equation}
\lambda \rightarrow -\lambda-(d-1). \label{involution}
\end{equation}
(see Eq. (\ref{mass})).  Therefore for all $\xi \in C^+$ the waves $\psi_\lambda(x, \xi)$
and $\psi_{-\lambda-d+1}(x, \xi)$ all have the same mass as given in Eq. $(\ref{mass})$.

The squared mass is real and positive for
\begin{enumerate}
\item \begin{equation}\lambda =-\frac{d-1}2 - i\nu \ \ \ (\nu \in {\bf R})\ \ \rightarrow  \ \ m^2 =   \frac{(d-1)^2}{4R^2} +\frac{\alpha^2}{ R^2}
\label{principal}\end{equation}
In a group-theoretical language  this case corresponds to the principal series of representation of the de Sitter group. In this case the involution (\ref{involution})
acts as the complex conjugation
$$ \lambda \to -\lambda - (d-1) = -\frac{d-1}2 + i\nu = \lambda^* $$
\item
 \begin{equation}\lambda = -\frac{d-1}2 - \alpha,\ \ \ \left(\alpha \in {\bf R}, \ \ \ |\alpha| < \frac {d-1}2\right)\ \ \rightarrow \ \ m^2 =   \frac{(d-1)^2}{4R^2} -\frac{\alpha^2}{ R^2}\label{complementary}\end{equation}
This case correspond to the complementary series of representation.
Now $\lambda$ is real and the involution is no more the complex conjugation but it is the real reflection $\alpha \to - \alpha$. Complementary plane waves are real and do not oscillate. This fact is the source of the very bad infrared behavior of the quantum field theories in the complementary series.
In a idealized de Sitter phase at the inflation epoch all the particles of the standard model would have masses belonging to the complementary series. While with today's value of the cosmological constant all those particle have masses deep inside the much safer principal series. One should be aware of those infrared divergences at the inflation era which put many field theoretical computation in danger (even if there is always a cheap way out: invoking a {\em quasi} de Sitter phase - which is good for other purposes - to avoid facing all such problems).
\item For real $\lambda >0$ or $\lambda<-(d-1)$ the squared mass is real but negative.
\end{enumerate}

The problem with the waves (\ref{planewaves}) is that they are not globally
well defined; the only notable exception is when $\lambda = n \in {\Bbb N}$;
here the waves are polynomials in the ambient spacetime coordinates.
A given lightlike forward vector $\xi \in C^+$ splits the de Sitter hyperboloid
into two halves separated by the hyperplane $\xi \cdot x = 0$.
To glue together the waves defined in the above halves of the de Sitter spacetime one needs to specify
the coefficients $a(\lambda)$ and $b(\lambda)$ to define the (singular) global waves as follows:
\begin{equation}
\psi^{(a,b)}_\lambda(x, \xi) =  a(\lambda) \theta (\xi \cdot x) (\xi \cdot x)^\lambda + b(\lambda) \theta (-\xi \cdot x) |\xi \cdot x|^\lambda.
\label{planewavesab}
\end{equation}
The coefficients $a(\lambda)$ and $b(\lambda)$ actually encode the physical properties
of the vacuum of the corresponding QFT.
In particular, the so called Bunch-Davies or Euclidean vacuum
\cite{N}
\cite{Th}
\cite{GiH}
\cite{ChT}
\cite{BuD}
\cite{Allen}
\cite{bmg}
\cite{bros}
which plays an extremely important role in the physical applications of de Sitter QFT,
is related to very precise analyticity properties of the plane waves which uniquely determine the coefficients $a(\lambda)$ and $b(\lambda)$,
apart from a global normalization that is further determined by the Canonical Commutation Relations.

The construction goes as follows \cite{bmg}
\cite{bros}: let us introduce the complexified de Sitter
space-time
\begin{equation}
X^{(c)}_{d}= \{z = x+iy \in {{\bf C}}^{d+1}:
{z_{0}}^{2}-{z_{1}}^{2}-\ldots- {z_{d}}^{2}=-R^ {2}\}.
\label{hypec}
\end{equation}
The
physically relevant global waves can be defined as analytic
functions for $z$ in the tubular domains $ {\cal T}^{+}$ or $
{\mathcal T}^{-}$ of $X^{(c)}_{d}$:
\begin{eqnarray}
{\cal T}^{+} = \left({\bf R}^{d+1} + i {V^{+}}\right)\cap
X^{(c)}_{d} = \{z = x+iy \in X^{(c)}_{d}:
y^2 > 0,\;y_0 >0\},
\cr{\cal T}^{-}= \left({\bf R}^{d+1} - i {V^{+}}\right)\cap
X^{(c)}_{d} = \{z = x+iy \in X^{(c)}_{d}:
y^2 > 0,\;y_0 < 0\},
\label{tubi1}
\end{eqnarray}
where $ T^\pm = {\bf R}^{d+1} \pm i {V^{+}}$ are the forward and
backward tubes in the ambient complex Minkowski space ${{\bf
C}}^{d+1}$.
For $z\in
{\cal T}^{+}$ or $ z \in {\cal T}^{-}$, $\xi \in C^+$ and $\lambda \in {\bf C}$ we then define
\begin{equation}
{\psi}_{\lambda} (z ,\xi )= \left( { {\xi}\cdot z}
\right)^{\lambda }. \label{dSwaves}
\end{equation}
 The phase
is chosen to be zero when the argument is real and positive.

In Minkowski space, the physical meaning of such
holomorphy properties is to provide the positivity of the
energy-momentum spectrum in the Hilbert space of the theory \cite{sw}.
In the de Sitter case they give rise to the thermal interpretation for the Bunch-Davies vacua \cite{bros}.
The above holomorphic waves provide us with the following choice of the coefficients $a(\lambda)$ and $b(\lambda)$:
\begin{equation}
\psi^{(\pm)}_\lambda(x, \xi) =  \theta (\xi \cdot x) (\xi \cdot x)^\lambda + e^{\pm i \pi \lambda} \theta (-\xi \cdot x) |\xi \cdot x|^\lambda.
\label{planewavesab}
\end{equation}

\section{Normalization (Tachyons).}
The waves $\psi^{(\pm)}_\lambda(x, \xi)$ are yet unnormalized.
We now proceed to evaluate the overall normalization  by the method of
canonical quantization.
Recall that, given two solutions of the Klein-Gordon equation, their Klein-Gordon product is defined by
\begin{equation}
(f,g) = i \int_\Sigma n^\mu(f^* \partial_\mu g -g\, \partial_\mu f^*  )  \sqrt{h}\,  d^{d-1}x, \label{kgproduct}
\end{equation}
where $h_{ij}$ is the induced metric on a arbitrary spacelike
surface $\Sigma$ and  $n^{\mu}$ is the forward directed unit vector normal to $\Sigma$.
First, we introduce an involution that generalizes the
complex conjugation and works for all complex $\lambda$ (and this is already a difference w.r.t. the Minkowski case). Define
$$ f_{\lambda,\xi}(x) = \psi^{(-)}_{-\lambda-d+1}(x, \xi),   \ \ \ \ f^*_{\lambda,\xi}(x) = \psi^{(+)}_\lambda(x, \xi). $$
After some effort, evaluation of the above integral gives
\begin{equation}
(f_{\lambda, \xi},f_{\lambda,\xi'}) =
 {2^{d+1}\pi^d R^{d-1}
e^{i\pi\left ( \lambda+{d-1\over 2} \right )}\over   {\Gamma(-\lambda)\Gamma(\lambda+d-1)}} \delta(\xi-\xi')= \frac 1 {c(\lambda)}\delta(\xi-\xi').
\label{r.4}
\end{equation}
Something special happens for the negative squared masses
relative to integer values $\lambda = n$: the corresponding plane waves are all non-normalizable zero modes.
The point is that such modes are homogenous polynomials  of degree $n$, globally defined and analytic in the complex de Sitter manifold (and actually in the whole complex ambient spacetime).
The space of such polynomials is finite dimensional and, as such, not rich enough to reconstruct
the infinitely many degrees of freedom of a quantum field.

Canonical quantization plus the above analyticity requirement for the modes uniquely fix the two-point functions as follows:
\begin{equation}
W_{(\lambda)}(x_1,\ x_2) = c(\lambda)\int_\gamma f_{\lambda,\xi}(x_1) f^*_{\lambda,\xi}(x_2)
d\mu_\gamma(\xi).
\label{tp}
\end{equation}
where $\gamma$ is  the sphere $\{\xi\in C^+ :\ \xi^0 = 1\}$ (or any ($d-1$)-cycle in $C^+$ homologous to it)
and $d\mu_\gamma(\xi)$ is the rotation invariant measure. To evaluate the two-point function we move as before to complex
events $z_1 \in {\cal T^-}$ and $z_2 \in {\cal T^+}$. The result is conveniently written in the following way
\begin{eqnarray} && w_{\lambda}(\zeta)  =
\Gamma(-\lambda) \; G_\lambda(\zeta)\ ,\ \ \ \zeta= z_1\cdot
z_2\,,\cr && \cr && G_\lambda(\zeta) = {\Gamma(\lambda+d-1)\over
(4\pi)^{d/2}\Gamma\left({d\over 2}\right)}\, {}_2F_1\left(-\lambda,\
\lambda+d-1;\ {d\over 2};\ {1-\zeta\over 2}\right )\ .
\label{r.1}\end{eqnarray}

\begin{enumerate}
\item Theories corresponding to $m^2>0$ (and the limiting massless minimally
coupled case $m^2=0$) have been widely studied in the literature
with important applications to inflation. Their
physical interpretation is however still under active discussion
\cite{polyakov,bem,polyakov2,Krotov,Akhmedov,Akhmedov2,bempoinc,bem2}. Furthermore there are huge differences between theories of the principal series (for which $ \ \  2 m R > d-1$) and theories of the complementary
series. The latter are much more difficult to use and interpret and have more exotic properties, and yet they are the most relevant for inflationary calculations.

\item Theories corresponding to negative or, more
generally, complex squared masses give rise to quantum fields which
are local and de Sitter invariant but they neither satisfy the
positive-definiteness condition nor admit a de Sitter invariant positive subspace of the associated linear space of local states.
At the moment, there is no acceptable physical interpretation for the
field theories in this family.

\item
The "would be" two point function
 corresponding to nonnegative integer
$ \lambda = n $ (the $n=0$ case being the massless
minimally coupled field) is simply
infinite because of the pole of
$\Gamma(-\lambda)$ at $ \lambda =n $ in the RHS of Eq. (\ref{r.1}). Note also that
\begin{equation}
_2F_1\left(-n,\ n+d-1;\ {d\over 2};\ {1-\zeta\over 2}\right ) =
\frac{\Gamma(n+1)\Gamma(d-1)}{\Gamma(n+d-1)}
C^{\frac{d-1}2}_n(\zeta)
\end{equation}
reduces to a Gegenbauer polynomial of degree $n$ (see \cite{HTF1});
it is a
holomorphic function without cuts and consequently with no quantum
feature.
\end{enumerate}

If we now examine the commutator (which indeed in QFT comes before the two-point function, see e.g. the discussion in  \cite{ms}) we might be surprised in seeing that
\begin{equation}
C_\lambda(x_1,x_2) =  W_\lambda(x_1,x_2)- W_\lambda(x_2,x_1)\label{kgcomm}
\end{equation}
exist and furthermore is local (i.e. vanishes at spacelike separation)  and de Sitter invariant
{\em for all complex $\lambda$,} including $\lambda= n$.
We therefore come to the conclusion that, notwithstanding (and actually thanks to) the infrared divergence of the two-point function $W_n(x_1,x_2)$,
a quantum field theory might and indeed do exist \cite{yaa} for the following integer values of the (negative) squared mass:
\begin{equation}
m^2 =  -\frac{1}{R^2} n(n+d-1).
\label{massn}
\end{equation}
Such fields are the de Sitter tachyons. They are local, covariant and admit a de Sitter invariant physical space.
But they have no Minkowskian counterpart. Their very existence contradicts the idea that de Sitter QFT is {\em locally} indistinguishable from Minkowski QFT and the eventual differences are only global. Tachyons quantum fields do exist because of the global properties of the de Sitter universe but their existence can be seen locally and they cannot be approximated by any Minkowskian quantum field theory.

\section{QFT: Tachyons}
The word  tachyon has a twofold meaning in (theoretical) physics:
it denotes either a would-be  particle traveling faster than light or a particle having a negative squared mass.

Despite the fact that tachyons are believed to be unphysical,
they play an important if not crucial role in many circumstances (as for instance in bosonic string theory)
and the question about their existence and physical meaning
is far from being settled.
The commonly accepted interpretation is that the appearance of
tachyonic degrees of freedom in a theoretical model points towards an instability
of the model, and this instability should be
described by tachyonic fields as opposed to real faster-than-light particles; the instability
can be treated by invoking the so-called tachyon condensation.

From a general point of view, the standard requirements of local commutativity, covariance and positive definiteness \cite{sw}
have proven to be incompatible for tachyonic quantum
fields on the Minkowski spacetime. Feinberg \cite{feinberg} considers tachyons as
scalar fermions abandoning the local commutativity property while Schroer \cite{schroer}
constructs local and covariant tachyon fields which
neither satisfy the positive-definiteness condition nor have a Poincar\'e invariant physical subspace direct quantum
mechanical interpretation.

In the de Sitter case the infrared divergence  of $W_\lambda$ at $\lambda = n$ opens the possibility
for tachyonic fields satisfying locality, covariance and positive definiteness {\em \`a la}  Gupta-Bleuler.

The situation is in a way similar (but in another way worse) to that of the two-dimensional massless
boson field and everyone knows the fundamental role that such a field play in conformal field theory.
Also in that case the infrared divergences render the axioms incompatible
and the way out is to quantize the field with indefinite metric.
There exists however a Poincar\'e invariant subspace identified
by a Gupta-Bleuler supplementary condition.

Following this analogy we introduce an infrared renormalized two-point function by substracting the diverging part:
\begin{equation}
\wh W_n(z_1,z_2)= \wh w_{n}(\zeta) = \lim_{\lambda\to n} \Gamma(-\lambda) \;
\left[G_\lambda(\zeta)-G_n(\zeta)\right] \end{equation}
Because $G_n(\zeta)$
has no discontinuity, the commutator $C_\lambda$ associated to
$W_\lambda$ (see (\ref{kgcomm})) tends to a well-defined limit as
$\lambda \rightarrow n$, without needing any subtraction, and this
limit is precisely the commutator associated to $\wh W_n$:
\begin{equation}
C_n(x,y) =  \lim_{\lambda\to n} C_\lambda(x,y) =  \wh W_{n}(x,y)-\wh W_{n}(y,x) \label{ren}.
\end{equation}
The price that has been payed in this construction is seen by reconstructing the pseudo-Fock space of the theory and the field operator. One sees that
\begin{equation}
\left[\Box  - \frac{1}{R^2}n(n+d-1)\right]\phi =  Q_n
\end{equation}
A local and de Sitter invariant quantization
of this family of field theories
(including the massless minimally coupled scalar) is thus possible at the
expense of an anomaly in the field equation.
Of course we are really interested in (tachyonic) fields satisfying the correct
field equation with no non-homogeneous term.
We are therefore tempted to impose  the condition
\begin{equation}
Q_n^- \Psi = 0 \label{supp}
\end{equation}
on physical states, where $Q_n^-$ denotes the "annihilation" part of the operator $Q_n$.
An extraordinary property is that the above condition also selects a
{\em positive} and {\em de Sitter invariant} subspace of the space of
local states
and therefore opens the way for an acceptable quantum mechanical
interpretation of the  de Sitter tachyons.

At the one particle level the supplementary condition can be simply written
\beq
\EE_n = \left\{ \Psi \in \CC_0^\infty(X_d)\ :
\int (x \cdot \xi)^{n} \ \Psi(x) dx= 0\right\} \label{rr}
\endq
where $dx$ is the de Sitter invariant measure on
the hyperboloid $X_d$, i.e. physical one particle must be annihilated by the zero modes.

Condition (\ref{rr})  selects a manifestly de Sitter
invariant subspace of the one-particle space.
The positive-definiteness of this subspace is a consequence of the following Fourier-type
(i.e. a "momentum space") representation of the renormalized two-point function, valid on $\EE_n\times \EE_n$:
\beq \wh W_n(z_1,z_2) = a_n \int_\gamma\int_\gamma (z_1\cdot\xi)^{1-d-n}\,
(\xi\cdot \xi')^n\,\log(\xi\cdot \xi')\,
(z_2\cdot\xi')^{1-d-n}\,d\mu(\xi)d\mu(\xi') \label{physical}
\endq
where $a_n$ is fixed by the CCR's.The above representation is valid for any pair of complex de Sitter events $(z_1,z_2)$ such that
$z_1$ belongs to the past tube $ \TT_-$ and $z_2$ to the future tube $ \TT_+$.
The two-point function $\wh W_n(x_1,x_2)$ is the boundary value on the reals from these tubes.

Two final remarks are in order. First, local and
covariant scalar tachyon quantum fields exist on a de Sitter universe for all  the masses
specified in Eq. (\ref{massn}). Why the allowed masses belong to a discrete set? This is because of
the existence of closed spacelike geodesics. Second, nothing similar exist in Minkowski spacetime.
What happens to the de Sitter tachyons in the flat limit?
A flat coalescence limit of the de Sitter tachyons can indeed be taken but the positive invariant subspace disappears in this limit
and one gets back to the impossibility of reconciling locality covariance and positive-definiteness.
The question whether one may formulate interacting theories including de Sitter tachyons remains entirely open.

\section{More about the complementary series}
Here we present another phenomenon
that exists for de Sitter  scalar quantum field theories having a
(positive) mass less or equal than $(d-1)/2R$ (i.e. for the complementary series).

In quantum field theory, a "particle" of mass $m_1$ is a quantum state
$$ \Psi^{(1)}_f=\left[\int \phi_{m_1}(x ) f(x) dx\right] \Psi_0 $$
obtained
by applying the smeared field operator to the vacuum.

The unexpected fact can be described is the following: when $ 2 m_1 R< (d-1)$ the same quantum state
can be alternatively seen as a composite made by two particles of certain mass $m_2 < m_1$. Furthermore, and perhaps more surprisingly, the same particle can also be seen
as a $n$-particle composite of fields with suitably low masses.
One would be tempted to call such a state a {\em bound} state even if there is no interaction term in the Lagrangian: the minimal coupling
of the scalar field to the de Sitter geometry is enough to produce this phenomenon. It is an infrared phenomenon: it does not take place for particles of sufficiently high mass.

To explain this astonishing result let us consider the Wick powers of the field $\phi(x)$.
In the simplest case, the two point function of the Wick squared field $:\phi^2:(x)$
\begin{equation}
\langle \Psi_0 :\phi^2:(x) :\phi^2:(y) \Psi_0\rangle = W_\lambda^2 (x,y)
\end{equation}
can be evaluated in the form of a K\"allen-Lehmann integral representation. Suppose first that
$\lambda$ is in the principal series (\ref{principal}). We get \cite{bempoinc}
\begin{eqnarray}
&& [w_{-\frac{d-1}2 + i\nu} (\zeta)]^2 = \int_{-\infty}^\infty
\kappa\,\rho(\kappa,\nu,\nu)\,w_{-\frac{d-1}2 + i\kappa}(\zeta)\,d\kappa\ ,
\label{r.31}\cr
&& {\kappa\,\rho(\kappa,\nu,\nu) = { { \left|\Gamma\left (\mu+{i\kappa \over 2}
\right ) \right|^2 \prod_{\epsilon, \epsilon' = \pm 1} \Gamma\left (\mu+{i\epsilon\kappa
\over 2} + i\epsilon'\nu  \right )}\over 2^{d+2}\pi^{d+1\over
2}R^{d-2}\,\Gamma(i\kappa)\Gamma(-i\kappa) \Gamma\left ({1\over
2}+\mu+{i\kappa\over 2} \right ) \Gamma\left ({1\over 2}+\mu-{i\kappa\over
2} \right ) \Gamma(2\mu)}}\cr &&
\label{r.30.1}\end{eqnarray}
where $\mu = {d-1\over 4}$; for real $\nu$ and $\kappa\not= 0$ is strictly positive the weight $\kappa\,\rho(\kappa;\nu,\nu)$ as it should.

The above formula in particular implies that, in
the presence of a suitable interaction term, any
``principal'' particle can decay into any pair of equal-mass
``principal'' particles violating the standard mass subadditivity principle (i.e. here a particle can decay into two heavier ones
and this is another feature that Minkowski QFT's do not share \cite{bem,bempoinc,bem2}).

Let us rewrite (\ref{r.31}) in terms of the hypergeometric function:
\begin{eqnarray}
\left[w_{-\frac{d-1}2 + i\nu} (\zeta)\right]^2 &=&
\int_{\bR} {\k\,\sh(\pi\k) \left|\Gamma \left (\mu+{i\k\over 2} \right)\right|^4\,
\prod_{\epsilon, \epsilon' =\pm 1}
\Gamma \left ( \mu +{i\epsilon\k\over 2}+{i\epsilon'\nu} \right )\,\over
2^{d+5}\pi^{d+{5\over 2}}\Gamma\left({d\over 2}\right)
\Gamma\left({d-1\over 2}\right)R^{2d-4}}\,
\ \times\cr
&\times &
F \left (
{d-1\over 2}+i\k,\ {d-1\over 2}-i\k\ ;\ {d\over 2}\ ;\
{1-\zeta \over 2} \right ) d\k\ .
\label{c.1}\end{eqnarray}
We have already described the poles of the lhs.
As regards the rhs, the integrand is meromorphic in $\k$ and $\nu$ and
has no singularity when both are real.
We may analytically continue the integral in the complex variable $\nu$; let us  suppose that
$$\Re \nu >0, \ \ \  \alpha = \Im \nu >0.$$
The poles of the functions $\k \mapsto
\Gamma(\mu \pm i\k/2)$ are at $\k = \pm 2i(\mu+k)$ ($k \ge 0$
integer), and are independent of $\nu$. The other poles of the
integrand are given by ($k \ge 0$ integer):
\begin{equation}
\pm{ i\k \over 2}+\mu \pm i\nu+k = 0
\label{c.3}\end{equation}

The poles $\k -2i(\mu +i\nu +k) =0$  and the poles $\k
+2i(\mu -i\nu +k) =0$  are on the line $-2\Re \nu
+i\bR$. Their mutual distances do not change as $\nu$ varies, and
they all move down as $\alpha$ increases. The poles $\k +2i(\mu
+i\nu +k) =0$ and $\k -2i(\mu -i\nu +k) =0$ are the opposites of those described before.
They lie on $2\Re \nu +i\bR$ and move up as $\alpha$ increases.
When $0< \alpha< \mu$,
no pole reaches the real axis and
the formula (\ref{r.31}) continues to hold.
This is true in particular for purely imaginary $\nu = i\alpha$
and therefore  formula (\ref{r.31}) is valid for the principal series and
half of the complementary series,  i.e for all the masses satisfying
$$m^2 > \frac{3(d-1)^2}{16R^2}.$$
When $\alpha$ reaches the threshold $\mu$ Eq. (\ref{r.31}) has to be replaced by a contour integral.
For $\mu < \alpha  < \mu+1$ we can extract the
residues of the poles at $\k = \pm 2i(\mu +i\nu)$. A similar
situation occurs when the successive poles $\k = \pm 2i(\mu
+i\nu+k)$ cross the real axis.
For $\Re\nu >0$, $\alpha \ge 0$, $\alpha - \mu \notin \bZ$,
$N = \max \left \{j \in \bZ\ :\ j < \Im \nu - \mu\right \}$  Eq. (\ref{r.31}) is modified by the appearance of a sum of discrete contributions:
\begin{equation}
\left[w_{-\frac{d-1}2 + i\nu} (\zeta)\right]^2 = \int_\bR
\k\,\rho(\k, \nu, \nu)\,w_{-\frac {d-1}2 + i\k}(\zeta)\,d\k\ + \sum_{k=0}^{N}
A_k(\nu)\,w_{-\frac {d-1}2 -2(\mu +i\nu+k)}(\zeta).
\label{c.7}\end{equation}
When $\nu$ tend to $i\alpha$  the above formula will continue to hold provided both parts of the rhs
remain meaningful. Therefore, if $0< \alpha < (d-1)/2$,
$\alpha - \mu \notin \bZ$, and
$N = \max \left \{j \in \bZ\ :\ j < \alpha - \mu\right \}$,
\begin{equation}
\left[w_{-\frac{d-1}2 + \alpha} (\zeta)\right]^2 = \int_\bR \k\rho(\k,i\alpha,
i\alpha)\,w_{-\frac{d-1}2+ i\k}(\zeta)\,d\k + \sum_{k=0}^{N}
A_k(i\alpha)\,w_{2(k-\alpha)}(\zeta)\ .
\label{c.8.1}\end{equation}
The positivity of the K\"allen-Lehmann weight
\begin{eqnarray}
{ \k\rho(\k,i\alpha,i\alpha)} ={\left|\Gamma(\mu +{i\k \over 2}-\alpha)\right|^2
\left|\Gamma(\mu +{i\k \over 2}+\alpha)\right|^2
\left|\Gamma(\mu +{i\k \over 2})\right|^2\over 2^{d+2} \pi^{1+d
\over 2} R^{d-2}\Gamma(2\mu)
\left|\Gamma(i\k)\right|^2
\left|\Gamma(\mu +{1\over 2}+{i\k \over 2})\right|^2}
\label{c.9} \end{eqnarray}
is obvious. A little work shows that also coefficients
$A_n(i\alpha)$ are positive:
\begin{eqnarray}
{
A_k(i\alpha) = {
\frac{(-1)^k}{k!}{\Gamma(2\mu-2\alpha+k)\over \Gamma(2\mu-2\alpha+2k)} {\Gamma(2\alpha-k)\Gamma(2\mu+k)\Gamma(\alpha-k)\Gamma(2\mu-\alpha+k)\over
\Gamma(2\alpha-2\mu-2k)\Gamma(\half+\alpha-k)\Gamma(\half+2\mu-\alpha+k)} \over 2^{d-1} \pi^{d-1\over 2}R^{d-2}\Gamma(2\mu)}
}>0
\end{eqnarray}
This result in particular says that in dimension $d=4$ in the spectrum of a de Sitter scalar quantum field of mass
$$ m  < \frac{3 \sqrt 3 }{4R}$$
there are also stable particles with mass
$$ m_2 =\frac 1R \sqrt{4 m^2+3 \sqrt{9-4 m^2}-9}$$
in the two-particle Hilbert subspace of the Fock space of the theory in sharp contrast with what happens in the Minkowski case (see e.g. the discussion in \cite{sw}) .

In general when
\begin{equation}
m< \frac{1}{2R} \sqrt{\frac{(d-1)^2 (2 n-1)}{n^2}}
\end{equation}
discrete states appear in the first $n$ finite-particle subspaces of the Fock space of the theory.

\section{Final remarks}
de Sitter quantum field theory is perhaps the most studied quantum field theory on curved spacetimes.
Nevertheless, there is still room for surprises
and source of unexpected features.
Here we have exhibited two phenomena that distinguish it sharply
from Minkowski quantum field theory and contradict the widespread
belief that at least locally de Sitter QFT is indistinguishable from Minkowski QFT.
These features may play a role in the application of de Sitter QFT to cosmology.
Their physical meaning is presently under investigation.



\bibliographystyle{aipproc}   


\end{document}

\IfFileExists{\jobname.bbl}{}
 {\typeout{}
  \typeout{******************************************}
  \typeout{** Please run "bibtex \jobname" to optain}
  \typeout{** the bibliography and then re-run LaTeX}
  \typeout{** twice to fix the references!}
  \typeout{******************************************}
  \typeout{}
 }

\end{document}


\endinput